# Modeling Single Electron Transfer in Si:P Double Quantum Dots

K. H. Lee[1], A. D. Greentree[2,3], J. P. Dinale[1], C. C. Escott[1], A. S. Dzurak[1] and R. G. Clark[2]

[1] Centre for Quantum Computer Technology, School of Electrical Engineering and Telecommunications, University of New South Wales, Sydney, NSW 2052, Australia
[2] Centre for Quantum Computer Technology, School of Physics, University of New South Wales, Sydney, NSW 2052, Australia
[3] Centre for Quantum Computer Technology, School of Physics, University of Melbourne, Victoria 3010, Australia

**Abstract**

Solid-state systems such as P donors in Si have considerable potential for realization of scalable quantum computation. Recent experimental work in this area has focused on implanted Si:P double quantum dots (DQDs) that represent a preliminary step towards the realization of single donor charge-based qubits. This paper focuses on the techniques involved in analyzing the charge transfer within such DQD devices and understanding the impact of fabrication parameters on this process. We show that misalignment between the buried dots and surface gates affects the charge transfer behavior and identify some of the challenges posed by reducing the size of the metallic dot to the few donor regime.

## 1. Introduction

Quantum computation offers important potential advantages over conventional computing [1][2]. Using quantum mechanical superposition and entanglement as new resources, quantum computers (QC) are expected to be more powerful than conventional computers. Certain algorithms have already been discovered which take advantage of this increased power to show significant speed-ups compared with classical algorithms, e.g. factoring large numbers [3] or searching a database [4]. Irrespective of such algorithms, atomic-scale quantum devices represent a fundamental limit of integrated circuit technology, and the ability to realize even *classical* algorithms with such devices is of considerable interest. Fundamental to the realization of a QC are a scalable physical system with well characterized quantum states (usually two level quantum systems, referred to as qubits – quantum bits) and a method to read out the states of the qubits [5].

Solid-state systems offer the prospect of a scalable QC and much progress has been made in this field [6], with coherent control and two qubit coupling having being demonstrated using a superconducting qubit [7]. QC architectures based on P donors in Si (Si:P) are also being actively pursued due to their compatibility with conventional Si metal-oxide semiconductor (MOS) fabrication technology [8][9]. In broad terms there are three main schemes for realizing donor based QC's, namely where the qubits are defined by nuclear spin [10], electron spin [11][12] and electron charge [13]. One of the most promising techniques for readout is via fast charge detection with highly sensitive electrometers – e.g. single electron transistors (SET) [14]. This may be for direct readout (e.g. the charge scheme in [13]) or for indirect readout via spin-to-charge conversion [10][11][15][16][17].

Recent work on Si:P qubits has focused on the development of test devices to investigate charge qubits using a 'top-down' fabrication approach; i.e. incorporation of P atoms by controlled ion implantation [18]. These double quantum dot (DQD) test devices were designed to demonstrate controlled electron transfer between buried dots via application of a differential bias to surface control gates. Details of electrical measurements on these devices will be presented elsewhere [19]. A scanning electron microscope image of a DQD device is presented in figure 1. Approximately 600 P ions are implanted at each of two sites (red circles in Fig. 1) to create two buried metallic dots. Ultimately the number of donors in each dot would be reduced to one in order to realize a charge qubit [13].

This paper focuses on modeling the electron transfer between the dots. The first half of the paper details the framework for analyzing the charge transfer behavior in a DQD device. This involved constructing an equivalent circuit model of the DQD device, with the capacitance between the various circuit features calculated using FASTCAP (a multipole accelerated 3-D capacitance extraction program [20][21]). This allowed quick calculations of the capacitance for complex shapes. In the second half of the paper we examine the effects of dot misalignment (with respect to the surface control gates) and the dot size on the charge transfer detected by the SET. This was the primary aim of our modeling studies. We also examine how this result could be applied to estimate the misalignment in experimentally measured devices. Whilst our discussion in this paper is limited to the DQD device architecture, the techniques presented in this paper are easily extendable to novel nano-scale devices based on all metallic systems, and hybrid metal-implant structures.

## 2. Device Architecture and Fabrication

We describe the device architecture and fabrication to show features of its operation and the computational model. Figure 1 shows the three main components in the present scheme: SETs, dots and control gates. SETs are employed as sensitive electrometers for qubit operation (readout of charge on the double-dot) [14][22] with noise rejection being achieved by cross-correlating the two SET signals. This is to distinguish between signals generated by the charge transfer between the dots, which is detected by both SETs, and signals generated by background noise which couples more strongly to only one SET [23]. For maximum sensitivity, the SETs are operated halfway up the Coulomb blockade oscillation peak where slight changes in potential leads to a large change in source-drain current $I_{SD}$.

The Si:P quantum dots are formed using phosphorus ion implantation with the implantation dose set to ensure that the resulting electron system is metallic, achieved by using a P doping density above the metal-insulator transition (MIT): $\sim 3.5 \times 10^{18}$ cm$^{-3}$ for Si:P [24]. Above this density the P atoms are packed densely enough to form extended states. Metallic dots are critical to ensure a large number of free electrons for transfer, together with equally spaced energy levels, so that periodic charge transfer could be observed. At densities close to the MIT we would have metallic P regions surrounded by insulating P regions, leading to a complex, multi-dot system. Modeling results indicated that dots separated by 100 nm (centre-to-centre) would lead to 40nm–diameter metallic regions with a barrier width of 60 nm. This would allow inter-dot charge transfer to be achieved with moderate gate biases, but far enough apart that the two islands remained as distinct dots separated by a barrier.

Charge transfer between the dots is controlled by barrier (B) and symmetry (S)-gates ($S_L$ and $S_R$). The B gate controls the height of the barrier in the double well potential, whilst applying biases between the $S_L$ and $S_R$ gates ($V_{SL}$, $V_{SR}$) changes the symmetry of the double well potential and allows for tunneling to occur between the dots. However $V_{SL}$ and $V_{SR}$ also induce charge on the SET island. In experimental measurements [19] and the modeling results presented here this is compensated for by applying a small bias on the SET gates to maintain a constant level of induced charge on the SET island. These compensation gates are labeled $g_1$ for SET1 and $g_2$ for SET2.

The DQD devices investigated by Buehler et al. [19] were fabricated on Si wafers with a high quality 5nm thick $SiO_2$ layer. Since the fabrication involved multiple lithography and metallization processes, Ti/Pt alignment makers defined by high-resolution electron beam lithography (EBL) ensured a common reference point. SET aerial, B, $S_L$ and $S_R$ gates were defined using EBL with the resulting Ti/Au metallization [25] producing features with 20 nm linewidth and 30 nm thickness. A subsequent process involving shadow evaporation [26] and in situ oxidation was used to create the Al/Al$_2$O$_3$/Al tunnel junctions required for the SETs. The thickness of the Al evaporation was ~30 nm. In-depth discussion on DQD device fabrication can be found in [8] and [18].

## 3. Charge Transfer Theory

To analyze the charge transfer in a double dot system, we construct an equivalent electrical circuit (figure 1). Using the equivalent circuit two parameters are calculated: *Charge transfer bias* and *SET island induced charge*. The former is necessary for direct comparison with experiment, while the latter is an important indicator of the experimentally attainable signal strength. These parameters are calculated using charge quantization on the metallic quantum dots and SET islands. For a review of the techniques involved see [27]. Furthermore this technique has been applied to analyze the electron transfer in a metal double dot test device and good agreement has been found between the modeling and experimental results [28].

*3.1 Charge Transfer Bias*

To calculate the charge transfer bias we assume the existence of a finite number of allowed charge configurations and calculate the energy of each configuration as a function of applied bias on the two S-gates. Charge transfer from one charge configuration to the next occurs when there is an energy degeneracy between the charge configurations and the system changes from one stable configuration to the next. This information is visualized as a plot showing the location of the degeneracy points for varying S-gate voltages (figure 2).

We make the following assumptions within our model to allow comparison with the experimental data. Firstly we assume that the SETs are compensated – this way we ignore the SET island charge for the purposes of calculating the transfer bias. This leads to a considerable simplification, reducing the number of charge configurations to be considered. Secondly we assume that the double-dot system has no source of electrons and must maintain the total charge (i.e. initially charge neutral). Therefore we arbitrarily assume that the total charge is zero and define the excess charge on dot $d_1$ ($d_2$) as $q_{d1}$ ($q_{d2}$). We label the allowed charge configurations in terms of the number of charges which have been transferred from $d_1$ to $d_2$, which we label $x$. The polarization of the double-dot system is therefore

$$P = (q_{d1} - q_{d2})/q_e = 2x/q_e \quad (1)$$

where $q_e$ is the electron charge. $x$ is allowed to vary over a range sufficient to ensure that a stable minimum is correctly identified for every point in the bias space.

To calculate a stable energy configuration, we follow the standard method. For every bias (defined by a vector of all applied biases, assuming $V_{S1} = V_{S2} = V_{D1} = V_{D2} = 0$), the gate induced charge is:

$$\tilde{Q} = \tilde{C} V$$

$$= \begin{bmatrix} C_{SLd1} & C_{SRd1} & C_{g1d1} & C_{g2d1} \\ C_{SLd2} & C_{SRd2} & C_{g1d2} & C_{g2d2} \end{bmatrix} \begin{bmatrix} V_{SL} \\ V_{SR} \\ V_{g1} \\ V_{g2} \end{bmatrix} \quad (2)$$

Hence elements of the $\tilde{C}$ matrix represent the coupling between the various gates and the dots. For example $C_{SLd1}$ is the capacitive coupling between $S_L$-gate and $d_1$ whilst $C_{g1d2}$ is the capacitive coupling between $g_1$ and $d_2$. The numerical values of the matrix elements were obtained from the FASTCAP modeling.

As we are interested in the charge transfer induced by $S_L$ and $S_R$ with compensated gates, $V_{g1}$ and $V_{g2}$ in equation 2 will be functions of $V_{SL}$ and $V_{SR}$. This dependence is determined by balancing the induced charge on the island due to all sources, and in general must be solved simultaneously for $V_{g1}$ and $V_{g2}$, however to good approximation we can ignore the cross coupling of one SET control gate to the opposite island, i.e. we set $C_{g1d2} = C_{g2d1} = 0$. In this case the compensation voltages become

$$V_{g1} = \frac{-(V_{SL} C_{SLd1} + V_{SR} C_{SRd1})}{C_{g1d1}} \quad (3)$$

$$V_{g2} = \frac{-(V_{SL} C_{SLd2} + V_{SR} C_{SRd2})}{C_{g2d2}} \quad (4)$$

which were the compensation biases applied in the following discussion.

We now consider a configuration of charge on the double dot, $[-x,x]^T$. The total charge $Q$ for this configuration is the sum of the gate induced charge $\tilde{Q}$ and the excess charge:

$$Q = \tilde{Q} - q_e \begin{bmatrix} q_{d1} \\ q_{d2} \end{bmatrix} \quad (5)$$

The energy of each charge configuration is

$$E_x = \frac{1}{2} Q_x^T C^{-1} Q_x \quad (6)$$

where the elements of $C$ represents either the coupling between the dots ($C_{d1d2}$) or the self capacitance ($C_{sum}^{d1}, C_{sum}^{d2}$)

$$C = \begin{bmatrix} C_{sum}^{d1} & -C_{d1d2} \\ -C_{d1d2} & C_{sum}^{d2} \end{bmatrix} \quad (7)$$

Domains of stable charge configurations are bounded by the lines of degeneracy, which can be solved for simply by numerically determining the biases which satisfy

$$Q_x^T C^{-1} Q_x = Q_{x+1}^T C^{-1} Q_{x+1} \quad (8)$$

### 3.2 SET Island Induced Charge

A charge transfer event on the double dot induces a charge in the SET island in a fashion analogous to an extra control gate. This change in induced charge $\Delta q$ is observed as a change in $I_{SD}$. For a given noise level in the SET $\Delta q$ sets the limits on measurement fidelity and readout time [22]; a large $\Delta q$ corresponds to an easily resolved signal, whereas if $\Delta q$ too small the change in $I_{SD}$ may be too small to practically resolve.

The method for determining $\Delta q$ is similar to that used above for the charge transfer bias, but in this case we concentrate on the charge configurations of the SET island, and calculate the SET charge transfer points for two separate polarizations of the double dot. We proceed as before, however this time assuming that SET2 is compensated, and it is the induced island charge of SET1 that we are interested in. The opposite case can be calculated trivially. We again calculate the induced island charge, however this time we take into account SET1.

$$\tilde{Q} = \tilde{C} V$$

$$= \begin{bmatrix} C_{SLd1} & C_{SRd1} & C_{g1d1} & C_{g2d1} \\ C_{SLd2} & C_{SRd2} & C_{g1d2} & C_{g2d2} \\ C_{SLi1} & C_{SRi1} & C_{g1i1} & C_{g2i1} \end{bmatrix} \begin{bmatrix} V_{SL} \\ V_{SR} \\ V_{g1} \\ V_{g2} \end{bmatrix} \quad (9)$$

The subscripts definitions are the same as for equation 2, with the addition of $i_1$ corresponding to SET1 island.

We now consider two sets of charge configurations, in each case the SET island charge ranging from [-y:y] and in one case the double dot configuration being [0,0] and the other being [-1,1]. The actual double dot configurations can be chosen arbitrarily, provided they differ by a single charge transfer event, as we are interested in the minimum induced charge due to a single electron transfer process on the double dot.

The capacitance matrix in this case is

$$\mathbf{C} = \begin{bmatrix} C_{d1}^{sum} & -C_{d1d2} & -C_{d1i1} \\ -C_{d1d2} & C_{d2}^{sum} & -C_{d2i1} \\ -C_{d1i1} & -C_{d2i1} & C_{i1}^{sum} \end{bmatrix} \qquad (10)$$

Graphs showing charge transfer points on the SET island for the [0,0] and [-1,1] dot configurations are shown in figure 3 as a function of $V_g$, with constant $V_{SL}$, $V_{SR}$ and $V_{g2}$. The distance between the charge transfer points is when SET1 gate has induced a change in the SET charge state by one electron. The extra fractional induced charge due to a change in the configuration of the double dot can then be seen graphically as a shift in the transfer pattern. The induced charge is a function of the coupling between the double-dot and the SET island, and does not depend on the magnitude of the gate coupling. Therefore the induced charge seen in each bias direction should be the same value, as is seen here with the induced charge being ~0.05*e*.

## 4. Computational Modeling

This section discusses the issues related to FASTCAP modeling; partitioning of the conductor surface and constructing the DQD device model.

*4.1 Multipole Expansion and Influence on Partitioning*

FASTCAP requires that the conductor surfaces be partitioned into *n* panels. The strategy for partitioning can be appreciated by understanding the multipole expansion algorithm employed in FASTCAP [20][21]. Fundamentally the capacitance between *n* conductors can be determined by raising the potential of one conductor to unity and grounding the remaining *n-1* conductors. Capacitance between the grounded and the unity potential conductor is then the total charge on the grounded conductor. This is known as 'direct evaluation' and has a computational cost of $n^2$.

Advantage offered by multipole expansion is realized when the panels are well separated – as it relies on approximating the far field contribution to the potential. FASTCAP has been implemented so that it uses multipole expansion only if the radius of the multipole region is less than half the distance between the centers [20]. This requirement influences the partitioning of the conductor surface. Finer partitioning (smaller panel sizes) allows for greater exploitation of the advantage offered by multipole expansion. However finer partitioning requires more panels to partition the conductor surface, leading to larger memory requirement and increased computation time. Hence constructing a FASTCAP model inescapably involves a compromise between speed and memory requirements.

*4.2 Model Construction*

A FASTCAP model of the DQD device was generated using the pattern files for EBL and incorporating the device features described in the fabrication process (e.g. thickness and linewidth). Hence, the resulting model was made to be geometrically accurate to the actual device and it consisted of approximately 49500 panels. The calculations were performed on the Australian Partnership for Advanced Computing (APAC) National Facility's 1GHz Compaq AlphaServer SC. The calculations required 3.4GB of memory and took on average 1.5 hours to perform.

In constructing the model, two further approximations were required. Firstly, the current release of FASTCAP does not support metal panels in contact with dielectric surfaces [29]. As a result, the bottom surfaces of the metal structures were separated from the top surface of the oxide layer by an air gap of 0.001 nm. Calculations using FASTCAP have shown that the error in the capacitance decreases as the size of the air gap decreases (figure 4). In addition the electric field visualization using FlexPDE has revealed the minimal effect of this air gap. The second approximation involved modeling the P dot as a conductor measuring 40x40x10 nm, this being the size of the metallic region according to a simple application of the MIT to the implanted region. This approximation afforded a great simplification, and is expected to be fairly good in the limit of high density implants with a large number of donors, or for P density close to the MIT.

*4.3 Comparison with Experimental Results*

The accuracy of the FASTCAP model was checked by comparing the calculated capacitance with experimentally measured values (table 1). Capacitance between a SET and gate can be measured by observing the periodicity of the Coulomb blockade oscillation [30]. This is due to the tunneling of a single electron through the SET island by applying a voltage on one of the gates (while leaving the others grounded). Periodicity of the oscillation *ΔV* is given by:

$$\Delta V = e/C_g \qquad (11)$$

where *e* is the electron charge and $C_g$ is the capacitance between the SET and the gate. Overall there is a good agreement between the measurement and calculation, given experimental variability and variations due to device fabrication.

|  | B Gate | $S_L$ Gate | $S_R$ Gate | SET1Gate |
|---|---|---|---|---|
| **Calculated** | 25.0 | 24.3 | 10.3 | 23.4 |
| **Measured** | 24.0 | 22.7 | 13.1 | 25.6 |
| **SD** | 3.35 | 2.44 | 2.17 | 2.45 |

Table 1. Calculated and measured capacitance between SET1 and a control gate. SD is the standard deviation encountered in the sample of measured devices. All capacitance are expressed in aF.

## 5. Modeling Results

*5.1 Effects of Dot Misalignment*

As previously mentioned Ti/Pt alignment markers are used as the reference point to align the nano-apertures (for ion implantation) and the subsequent EBL processes [18]. In an 'ideal' situation the buried dots could be located as in figure 1. However due to the accuracy of the EBL system (resolution and accuracy of the stage movements) this is rarely achieved in practice and the surface gates (fabricated after the ion implantation) are misaligned relative to the dots. The effects of misalignment were studied by displacing the location of the dot from the 'ideal' location by ±50 nm in the y direction ($\Delta y$) and ±90 nm in the x direction ($\Delta x$) with 10 nm step in both directions. This range of values corresponds to the approximate misalignment experienced in our EBL system. At each misalignment value the capacitance matrix was calculated using FASTCAP and this was used to determine the charge transfer by solving equations 1-8. Examples of the resulting charge transfer plots for various misalignment are presented in figure 5(a).

The middle plot in figure 5(a) represents the ideal case of perfect alignment between the dots and the surface gates. The lines corresponds to the charge transfer events – solutions to equation 8. Periodicity of the lines along $V_{SL}$ and $V_{SR}$ ($\Delta V_{SL}$, $\Delta V_{SR}$) are equal as $S_L$ and $S_R$ have equal effect on the charge transfer between the dots. In this configuration $C_{SLd1} = C_{SRd2} = 2.31$ aF and $C_{SLd1}/C_{SRd2} = 1$. However dot misalignment results in the asymmetric effect of the two control gates and the charge transfer plot is altered. Note that the asymmetry is a direct result of the extended structure of the gates. We use two parameters to characterize the charge transfer – periodicity and charge transfer angle $\theta$, which is defined as:

$$\theta = \tan^{-1}(\Delta V_{SR}/\Delta V_{SL}) \qquad (12)$$

Misalignment in the x direction varies the *relative* coupling of the $S_L$ and $S_R$ gate to the dot. For example, misalignment in the negative x direction (towards $S_L$ gate) results in the dots being more strongly under the influence of the $S_L$ gate. At $\Delta x = -50$ nm $\Delta y = 0$ nm (corresponding to left most plot in figure 5) $C_{SLd1} = 2.98$ aF, $C_{SRd2} = 1.54$ aF and $C_{SLd1}/C_{SRd2} = 1.93$. As a result $\Delta V_{SL}$ decreases while $\Delta V_{SR}$ increases, resulting in increased $\theta$. Misalignment in the y direction varies the *periodicity* of the charge transfer line as the *absolute* coupling of the $S_L$ and $S_R$ gate to the dot varies. For example misalignment in the positive y direction (away from the control gates) results in overall weaker coupling of the control gates to the dots. Hence, larger voltage changes are required for subsequent charge transfer events to occur and so $\Delta V_{SL}$ and $\Delta V_{SR}$ increases. At $\Delta x = 0$ nm $\Delta y = 50$ nm (corresponding to top-most plot in Figure 5) $C_{SLd1} = 1.45$ aF, $C_{SRd2} = 1.43$ aF and $C_{SLd1}/C_{SRd2} = 1.01$. It is important to note that this discussion is based on our DQD device architecture and any changes to the geometry (i.e. different placement of the gates) would lead to a different behavior.

Variation of $\theta$ and $\Delta V_{SL}$ over the misalignment region are presented in figure 6. These plots display the trend previously discussed. As seen in figure 6(b) minimum $\Delta V_{SL}$ is obtained when $\Delta x = -50$ nm. In this configuration the centre of the left dot centre is closest to the axis of the $S_L$ gate. In our region of interest the maximum coupling occurs at $\Delta x = -50$ nm $\Delta y = -50$ nm; corresponding to misalignment where the dot is directly underneath the $S_L$ gate. At this point $\Delta V_{SL} = V_{SL,min} = 47.5$ mV and $\theta = 82.6$ degrees. Over the misalignment region $\Delta V_{SL}$ varies over large dynamic range (47.5 mV to 8.36 V). To ensure that the variation in $V_{SL}$ across this range could be visualized the periodicity was expressed in terms of dB:

$$\Delta V_{SL} \text{ (dB)} = 20 \log_{10}(\Delta V_{SL}/V_{SL,min}) \qquad (13)$$

Results in figure 6 show that it is possible to estimate the misalignment between the dot and the surface gates by observing the charge transfer plot of a measured DQD device. figure 6(a) shows that for $\theta > 45$ degrees this corresponds to $\Delta x < 0$ nm and for $\theta < 45$ this corresponds to $\Delta x < 0$ nm. Furthermore figure 6(b) shows that small $\Delta V_{SL}$ (50-200 mV) corresponds to the dots being close to the $S_L$ gate. By extracting these characteristic features from the experimental charge transfer plot an estimate of the location of the dot could be determined. In other work, this technique has been successfully used to estimate the misalignment between an SET and a surface metal double dot [28].

### 5.2 SET Sensitivity Variation with Dot Size

The DQD serves as a prototype device to demonstrate the concept of a solid state implementation of a charge qubit. Ultimately such a device would consist of a single P atom, as opposed to a dot formed by tens or hundreds of P atoms. To reach this goal via top-down fabrication, the size of the P dot will be reduced, by decreasing the ion implantation dosage. Hence lower numbers of ions would pass through the aperture leading to fewer ions per dot.

We have modeled the effects of dot size on the signal produced by the SET. In reducing the dot size we have assumed that the metallic region varies isotropically with the number of ions per dot. That is, the aspect ratio of the metallic region is kept constant as the number of ions per dot varies. The inset in figure 7(a) shows the dimensions of the metallic region used for these studies. Subsequent modeling with Crystal-TRIM [31] and UT-MARLOWE [32] have confirmed that this a valid assumption – with the ratio of MIT variation in transverse (x-y) and lateral (z) directions with ion dose being similar.

To investigate the effects of dot size the dimension R was varied, with values: R= 10, 20,… 50 nm. For each dot size the capacitances between the dots and surface gates were calculated using FASTCAP. In these calculations there was no misalignment between the dot and the surface gates. Figure 7 shows that coupling capacitance between the respective SET and dot and $\Delta q$ decreases as the dot dimensions decrease. The results presented in figure 7(b) are important as they provide an important insight into the measurement time required by a RF-SET to achieve a high readout fidelity (error probability $P_e$ less than $10^{-6}$).

For a DQD device with 600 ions per dot the modeling predicts a $\Delta q \sim 0.05e$. At this $\Delta q$ the error probability model in [22] predicts that sub-microsecond measurement times are required to achieve a high readout fidelity. As $\Delta q$ decreases it becomes difficult to maintain a high readout fidelity, as the shift in the transfer pattern becomes increasingly difficult to discriminate. This is reflected in the error probability model as $P_e$ rapidly increases with decreasing $\Delta q$, since $\Delta q$ appears in the argument of an error function. Figure 7(b) show that for a dot with a MIT region measuring 10x10x2.5 nm (corresponding to a dot with tens of ions) $\Delta q \sim 0.02e$. Results from [22] suggest that signals smaller than $\Delta q \sim 0.01e$ cannot be resolved with high fidelity. Hence as we move towards the ultimate goal of two donor devices, more elaborate schemes may need to be devised to increase the capacitive coupling required to achieve high fidelity readout using 'single-shot' measurement such as the use of charge-shelving [33][34].

## 6. Conclusion

We have employed FASTCAP for an innovative application in nanoelectronics, beyond its original application of calculating the cross coupling capacitance between metal tracks in an integrated circuit device. Using FASTCAP, capacitances between the SETs, surface gates and buried Si:P metallic dots were determined and these were used to model charge transfer in a DQD. Results from these calculations have shown the effects of dot misalignment on the signal detected by the SET corresponding to the charge transfer – with misalignment in the x direction affecting the angle of the charge transfer and misalignment in the y direction affecting the periodicity. Using these results an estimation of the dot misalignment could be determined. Lastly, the effect of moving towards a smaller dot has been demonstrated with the decrease in $\Delta q$ presenting a challenge to achieving high fidelity readout using 'single-shot' measurement.

This study provides the basis of a useful tool in understanding the characteristics of a charge qubit device and the confirmation of metallic properties in Si:P double quantum dots. Modeling is important for interpretation of the experimental results and to allow further optimization of design parameters. More generally, the techniques involved are widely applicable in the construction of novel nano-scale devices based on all metallic systems, and hybrid metal-implant structures.

## Acknowledgement

This work was supported by the Australian Research Council, the Australian Government, the U.S. National Security Agency (NSA), the Advanced Research and Development Activity (ARDA) and the Army Research Office (ARO) under contract number DAAD19-01-1-0653. Authors thank the APAC National Facility for access to the computational facilities and D. Madina and D. J Reilly for computational support. We thank F. Green and V. Conrad for useful discussions regarding the modeling, V. Chan and D. J. Reilly for the device capacitance values, and T. Buehler, A.G. Ferguson, F.E. Stanley, A.R. Hamilton, D.N. Jamieson, C. Yang, C.I. Pakes for discussions on electrical measurements of DQD devices. K.H. Lee acknowledges the support of an Australian Postgraduate Award.

# Figure Captions

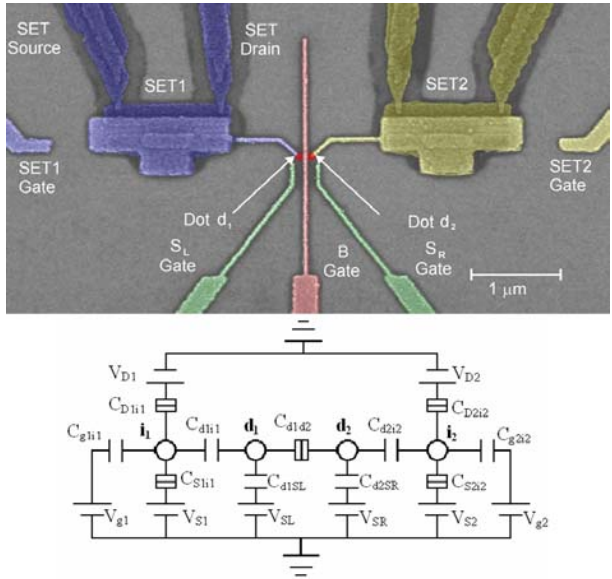

**Figure 1.** SEM image of the DQD with the SET, dot and control gates labeled (top). Equivalent electrical circuit of the DQD with the cross capacitances omitted for clarity (bottom).

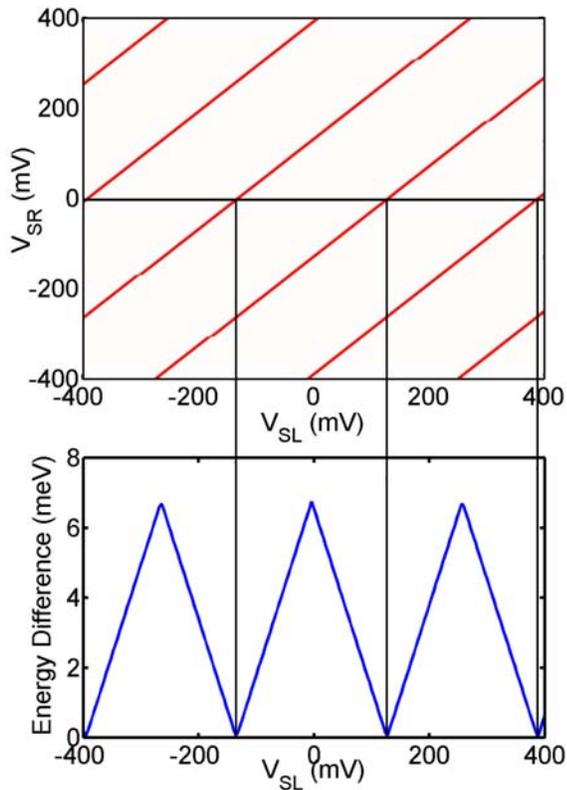

**Figure 2.** Plot showing the location of the degeneracy points for varying $S_L$ and $S_R$ gate voltages (a). Energy difference between the first and second states for $V_{SR} = 0$ mV and varying $V_{SL}$ (b). Charge transfer occurs when the energy difference between the two states is zero and this corresponds to the location of the lines in (a).

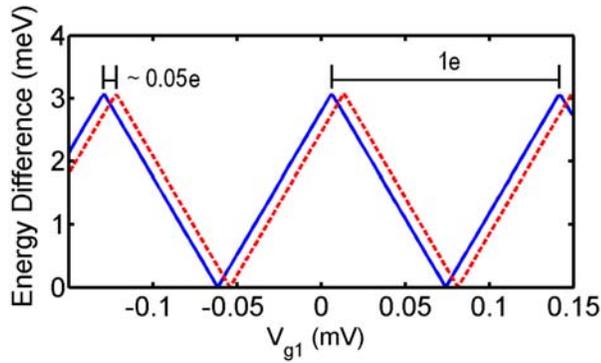

**Figure 3**. Charge transfer events on SET 1 as a function of $V_{g1}$ for different charge configurations of the double dot. The solid blue plot corresponds to the energy difference between the first and second states of varying SET charge for the double dot in the state [0,0]. The dashed red plot corresponds to the energy differences when the double dot is in the state [-1,1]. The shift in the transfer pattern is caused by the fractional induced shift due to a change in the double-dot configuration and is conventionally expressed as a fraction of the gate bias required to induce one whole electron difference on the SET. In this case the induced charge is ~0.05$e$.

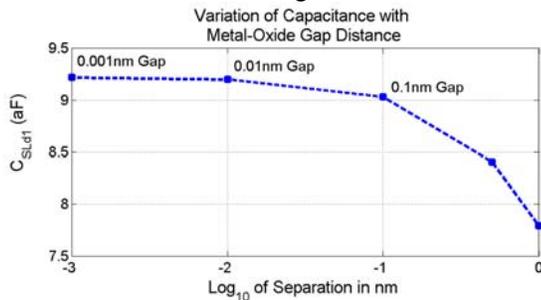

**Figure 4.** Effect of the air gap between the metal and oxide panels on the capacitance between $d_1$ and $S_L$ gate. Other capacitance values also display similar behavior. In this instance the dot was displaced from its ideal location (see Figure 5) by -50nm in the x direction and -120 nm in the y direction.

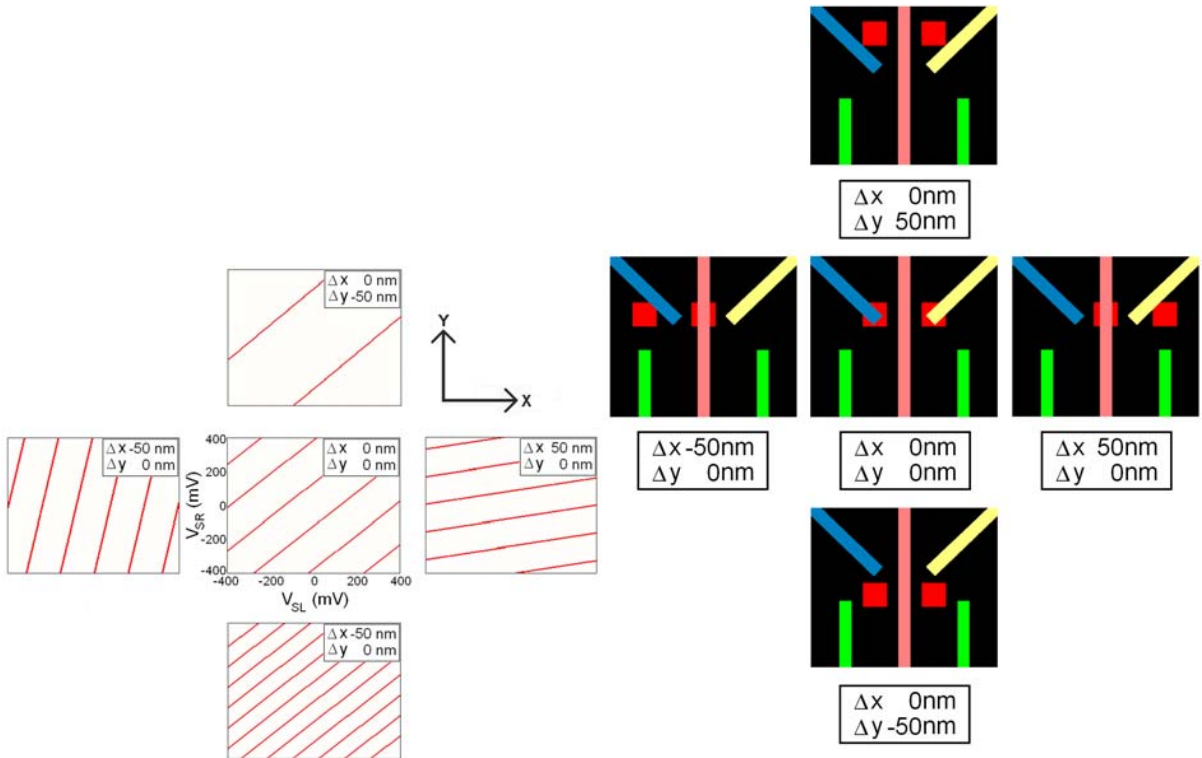

**Figure 5.** The effect of misalignment on the charge transfer (a). Schematic of the device central region showing the location of the dot for each of the misalignments studied (b).

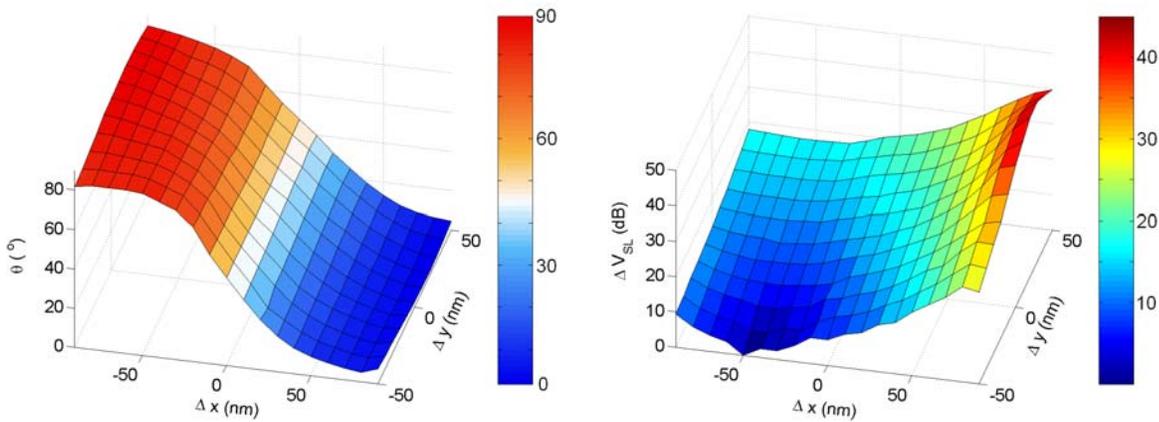

**Figure 6.** Variation of $\theta$ (a) and $\Delta V_{SL}$ (b) with the dot misalignment.

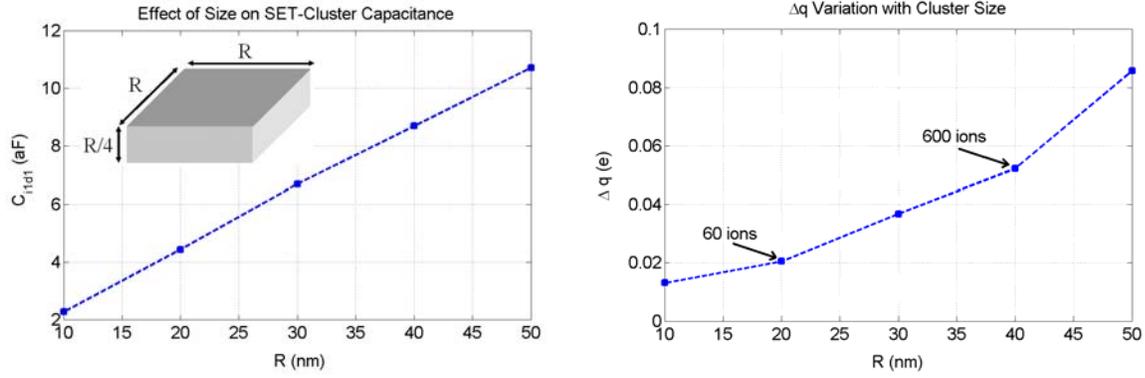

**Figure 7.** Effect of dot size on the Custer Device. Capacitive coupling between the respective SET and dot as a function of dot size (a). In this instance the capacitance between SET1 and d1 – $C_{i1d1}$. A more useful parameter is $\Delta q$ – SET sensitivity to detect the charge transfer between the dot (b).